\documentclass[12pt]{iopart}

\usepackage{graphicx}

\newcommand{\Bra}{\left\langle} 
\newcommand{\Ket}{\right\rangle}

\newcommand{\braket}[1]{\langle #1 \rangle}

\newcommand{\bubun}[2]{\frac{\partial #1}{\partial #2}}
\newcommand{\bibun}[2]{\frac{d #1}{d #2}}
\newcommand{\infinity}{\infty}

\begin{document}

\title[On the Distribution of State Values of Reproducing Cells]{ On the Distribution of
State Values of Reproducing Cells: the General Evolution Equation and its Applications }

\author{Katsuhiko Sato$^1$ and Kunihiko Kaneko$^{1,2}$}

\address{ $^1$ ERATO Complex Systems Biology Project, JST }

\address{ $^2$ Department of Pure and Applied Sciences, University of Tokyo, 3-8-1 Komaba,
Meguro-ku, Tokyo 153-8902, Japan }

\ead{sato@complex.c.u-tokyo.ac.jp and kaneko@complex.c.u-tokyo.ac.jp}

\begin{abstract}
Fluctuations of cell state, e.g., abundances of some proteins, have attracted much
attention both theoretically and experimentally.  The distribution of such state over
cells, however, is not only a result of intracellular stochastic process, but is also
influenced by the growth in cell numbers that depends on the state. By incorporating the
growth-death process into the standard Fokker--Planck equation for the probability
distribution, a nonlinear temporal evolution equation of distribution is obtained that
includes a self-consistent growth term.  The derived equation is generally solved
analytically by means of eigenfunction expansions.  By focusing on the case with linear
relaxation, two examples are considered as applications of the proposed general formalism.
First, by assuming that the growth rate of a cell increases linearly with the state value
$x$, the shift of the average state value $x$ due to the growth effect is shown to be
proportional to the variance of the state $x$ and the relaxation time, similarly with the
biological fluctuation- response relationship.  Second, when there is a gap in the growth
rate at some threshold value for the state $x$, existence of a critical gap value is
demonstrated, beyond which the average growth rate starts to increase. This critical value
is again obtained in terms of the relaxation time and the variance of $x$, all of which
are experimentally measurable quantities.  The relevance of the results to the analysis of
biological data on the distribution of cell states, as obtained for example by flow
cytometry, is discussed.
\end{abstract}






\maketitle

\section{Introduction}

Biological systems suffer fluctuations.  No intracellular biochemical process can avoid
fluctuations, because they arise from the motion and reaction of molecules.  For example,
gene expressions or abundance of some proteins in a cell fluctuate in time or by cells,
even if they are measured at the same time after a cell division, for cells with identical
genes (clones).  Indeed, Elowitz has explicitly measured the numeric fluctuations of
proteins in Escheria coli, by distinguishing intrinsic and extrinsic fluctuations
\cite{Elowitz}.  Such intracellular fluctuations have attracted both theoretical and
experimental attention \cite{Collins,Ueda,Oosawa,log-normal,Siggia,Elowitz2}, while the
significance of the phenotypic fluctuations for adaptation \cite{Kashiwagi} and evolution
\cite{Sato,Kaneko} has also been investigated.

In general, let us consider the fluctuation of some quantity $x$ characterizing the state
of a cell, such as the number of proteins or gene expressions.  Now, as a result of
intra-cellular dynamics, $x$ fluctuates among cells or in time.  Let us denote the
single-cell distribution of $x$ by $P_{single}(x)$.  In principle, it can be obtained by
repeating a single-cell measurement over an ensemble of cells.

Here, however, we must be careful about the choice of the initial ensemble itself for such
distributions.  The initial distribution of cells chosen for an experiment depends on
whether the cell can proliferate or not and the speed of cell replication, which may
depend on the cell state $x$.  Consider, for example, taking an ensemble of cells from a
culture.  Then the probability of choosing cells that have higher replication speeds will
be larger, and the initial distribution of $x$ will be biased accordingly.

This problem is prominent in the measurement of cells from continuous cultures using flow
cytometry or some other means \cite{Arnold}.  In flow cytometry, the characteristics of
each cell (e.g., the magnitude of fluorescence when a fluorescent protein gene is
introduced) are measured over a huge number of cells.  It is now established as a
standard, powerful tool to measure the distribution of states of cells.  Here, if the
growth rate of a cell is independent of the quantity $x$, the choice of cell ensemble is
not biased by the value $x$, and thus the observed distribution $P(x)$ by flow cytometry
is simply that given by the distribution $P_{single}(x)$.  On the other hand, if the
growth rate depends on $x$, the distribution $P(x)$ may be altered from the distribution
from single-cell dynamics.

As an illustration, consider the case in which $P_{single}(x)$ is a Gaussian distribution
around $x=x_0$, while the replication rate of a cell increases strongly with $x$ for
$x>x_0$, assuming that $x$ represents the abundance of some chemical that mediates the
growth of the cell.  In this case, it is naturally expected that the observed distribution
$P(x)$ should be biased towards $x>x_0$.

In general, the distribution $P_{single}(x,t)$ has been studied with the use of stochastic
processes to characterize the intra-cellular dynamics of the state $x$.  Established
mathematical tools such as Master's equation, Langeving's equation, and the Fokker--Planck
equation \cite{Haken,vanKampen} are applied for such studies.  On the other hand, as a
biological unit (cell) replicates, the number of cells increases accordingly.  This effect
of replication, then, must be incorporated with these stochastic processes, to include
both the single-cell fluctuations and the growth dynamics of the cells together.

Recently, there has been growing interest in exploring the relationship between the
fluctuations of intracellular state and the response of the state to the change in
external conditions, both theoretically and experimentally
\cite{Elowitz,Siggia,Paulsson,Arkin,Shibata-Fujimoto}.  For example, a change in the
concentration of some protein (or gene expression) against the change in the external
condition (e.g., concentration of some chemical in the medium) may be measured
experimentally, from which the response of such intracellular state to the environmental
change must be unveiled.  Here, however, the growth speed of a cell generally depends on
the intracellular state, e.g., the abundance of such protein, because the protein is
important to the function of the cell.  Hence, the measured change of the protein
concentration in response to external change involves both the internal change of the
intracellular state and the change in the cell number distribution caused by the
state-dependent growth rate.  Thus, we should develop a theoretical tool to distinguish
the two effects, based on the measurable quantities.  In the present paper, by setting up
an equation for $P(x,t)$ that takes into account both the intra-cellular stochastic
process and the state-dependent cell reproduction rate, we address this issue.

We first derive the evolution equation of the distribution $P(x,t)$ by extending the
Fokker--Planck equation to incorporate state-dependent growth.  (In the present paper,
`growth' means the replication of a cell, and the replication rate in time is called the
growth rate).  The derived equation includes a term for state-dependent growth, from which
is subtracted the average growth rate over all cells, leading to a source/sink term that
corresponds to the growth/death process of a cell.  The average growth rate gives a
self-consistent term that is nonlinear in distribution $P(x,t)$, but we can formally solve
the equation through the eigenvalue properties of a Sturm--Liouville-type operator.  After
giving a general formulation of the equation, we present two simple examples of this
formulation, by assuming the linear Langevin equation for the single-cell dynamics of the
state variable.  First, by considering the linear dependence of the growth speed on $x$,
we obtain a formula for the shift of the average value of the state $x$.  The shift is
proportional to the product of the variance of the state, the relaxation time, and the
proportion coefficient of the growth speed with $x$.  For our second example, we study the
case in which there is a threshold value of the state $x$ for growth, and derive a formula
for the change of $P(x,t)$ to `feel' the state-dependent growth.  Cautious remarks are
made on the interpretation of the distribution obtained from flow cytometry, while the
relevance of our theory to evolution is also briefly discussed.

Note that we do not discuss specific mechanism for the cell growth here.  Rather, we
introduce a function characterizing state-dependence growth generally and derive the
distribution function.

\section{Derivation of the equation for the distribution of cell state with
reproduction}

Let us first introduce a variable $x$, which represents a state value of a cell, for
example, a concentration of some chemical (or its deviation from the mean value).  We
assume that the temporal evolution of variable $x$ in a single cell obeys some Markovian
dynamics, that is, the value of $x$ at time $t$ is determined only by the value of $x$ at
{ some previous } time.  (Although biological systems may often retain some
memory, this assumption can be acceptable as a first-step approximation, and indeed is
adopted for most models.)  Based on this assumption, we consider the following Langevin
equation, which is often adopted:
\begin{equation} \frac{dx(t)}{dt} =-f(x(t))+\sqrt{g(x(t))} \;\eta(t) , \label{03}
\end{equation}
where $f$ and $g$ are functions of $x(t)$, that govern the dynamic behavior of the
variable $x$ (the function $g$ must be non-negative for all $x$); roughly speaking, the
function $f$ represents the force acting on the value toward its mean value and $g$
represents the strength of the diffusion at the value.  $\eta(t)$ is a Gaussian white
noise term having the statistical properties: $\langle \eta(t)\rangle = 0$ for any $t$ and
$\langle \eta(t_1)\eta(t_2) \rangle=2 \delta (t_1-t_2)$ for any $t_1$ and $t_2$.
{ The distribution function $P_{single}(x,t)$ indeed obeys the
Fokker--Planck equation
\footnote{ Here we have adopted Ito calculus; for Stratonovich calculus, one can simply
replace $f$ by $f- \frac{1}{2} \frac{d g}{dx}$ in equation (\ref{06}).  If $g$ is
constant, there is no difference.} derived from the Langevin equation (\ref{03})
\cite{Haken,vanKampen,kn:Risken}
\begin{equation}
\bubun{P_{single}(x,t)}{t} =\bubun{}{x} \left[ f(x) + \bubun{}{x} g(x) \right]
P_{single}(x,t).
\label{37} \end{equation}
}

{ We now introduce the growth (replication) of the cell, whose rate $\mu$
is dependent on the state value of $x$ in the cell, and is a function of $x(t)$, denoted
by $\mu(x(t))$. To derive the equation for the distribution $P(x,t)$ for this growth rate
of the cell, we first write down the change in the distribution function at time $t+\Delta
t$, given $P(x,t)$ at time $t$, as:
\begin{equation}
\hat{P}(x,t+\Delta t) = \int_{x_1}^{x_2} W (x,x',\Delta t) P(x',t) (1+\mu(x') \Delta t)
dx'
\label{18},
\end{equation}
} { where the term $(1+\mu(x') \Delta t)$ indicates the effect of cell
growth, while $W(x,x',\Delta t)$ is the transition probability that the system changes
from the state with $x'$ to that with $x$ during the time interval $\Delta t$, which is
determined by the Langevin equation (\ref{03}) uniquely.  Because of cell growth, the
distribution function $\hat{P}(x,t+\Delta t)$ obtained above is not normalized in general,
while the distribution $P(x,t)$ must be normalized.  To obtain the correct form of
$P(x,t+\Delta t)$, then, we must renormalize this distribution as:
\begin{eqnarray}
P(x,t+\Delta t) \!\!\!&=&\!\!\!  \frac{ \int_{x_1}^{x_2} dx' W (x,x',\Delta t) P(x',t)
(1+\mu(x') \Delta t) }{ \int_{x_1}^{x_2} dx \int_{x_1}^{x_2} dx' W (x,x',\Delta t) P(x',t)
(1+\mu(x') \Delta t) } \nonumber \\ \!\!\!&{\simeq}&\!\!\!   (\mu(x)-\bar{\mu}(t))
P(x,t) \Delta t + \int^{x_2}_{x_1} W (x,x',\Delta t) P(x',t) dx',
\label{38}
\end{eqnarray}
where we have used the property of the transition probability, $\int_{x_1}^{x_2}W
(x,x',\Delta t) dx=1$ for any $x'$ and any $\Delta t$, and retained in the second line
only the terms up to the first order in $\Delta t$. Here $\bar{\mu}$ is defined by
\begin{equation} \bar{\mu}(t)=\int_{x_1}^{x_2} \mu(x) P(x,t) dx
\label{33},
\end{equation}
which gives the mean growth rate of the cells at time $t$.  In equation (\ref{38}), taking
the limit $\Delta t \rightarrow 0$ and recalling the second term is reduced to the form of
equation (\ref{37}), we obtain
\begin{equation}
\bubun{P(x,t)}{t}=(\mu(x)-\bar{\mu}(t)) P(x,t) + \bubun{}{x} \left[ f(x)+ \bubun{}{x} g(x)
\right] P(x,t). \label{06}
\end{equation}
This is the equation we desired to derive, the time evolution equation for the
distribution function with $x$-dependent cell growth rate.  }

As in the standard Fokker--Planck equation for the probability, we take the no-flux
boundary condition as:
\begin{equation} \left. \left[ f(x)+ \bubun{}{x} g(x) \right] P(x,t)
\right|_{x=x_1,x_2}=0. \label{07} \end{equation}

If $\mu(x)=\mbox{constant}$, i.e., for $x$-independent cell growth, the first term in
equation (\ref{06}), $(\mu(x(t))-\bar{\mu}(t)) P(x,t)$, vanishes and accordingly equation
(\ref{06}) is reduced to just the usual Fokker--Planck equation (\ref{37}); the influence
of the state-dependent cell-growth appears only in the term $(\mu(x(t))-\bar{\mu}(t))
P(x,t)$, which plays the role of source (sink) in the distribution density, if the growth
rate at some point $x$ is greater (smaller) than the mean growth rate, $\bar\mu$.
Equation (\ref{06}) obtained above is nonlinear in $P$ because the term $\mu(t)$ involves
$P$ itself, so that it first looks rather difficult to analyze.  Fortunately, however, the
analysis turns out not to be so difficult, as will be shown in the next section.

\section{Analysis of the evolution equation of the distribution with growth} \label{15}

In this section, we examine the structure of equation  (\ref{06}), with the aid of linear
operators and eigenvalues.  We first introduce a linear operator \begin{equation}
L=\mu(x)+\bubun{}{x} \left[ f(x)+ \bubun{}{x} g(x) \right] \label{02}
\end{equation}
and rewrite equation  (\ref{06}) as
\begin{equation}
\bubun{P(x,t)}{t}=-\bar{\mu}(t) P(x,t)+L(x) P(x,t).
\label{39}
\end{equation}

As the operator $L$ is of the Sturm--Liouville type, we can, in principle, find all of its
eigenvalues and corresponding eigenfunctions, and all the eigenvalues are
real~\cite{kn:Morse}.  We denote the eigenvalues and the corresponding eigenfunctions by
$\lambda_i$ and $\phi_i(x)$, where the index $i$ runs over the non-negative integers,
$i=0,1,2,..$, and the eigenvalues are ordered so that $\lambda_i \geq \lambda_j$ for $i <
j$.  From the definition, $\lambda_i$ and $\phi_i(x)$ satisfy the relation
\begin{equation}
L(x)\phi_i(x)=\lambda_i \phi_i(x) \label{13}.
\end{equation}

In general, we can introduce the adjoint operator of $L$, denoted by $L^{\dagger}$, and
introduce the "left'' eigenfunctions of $L^{\dagger}$ denoted by $\psi_i(x)$, for the
eigenvalue $\lambda_i$.  As is well known, left and right eigenfunctions for different
eigenvalues are orthogonal and can be normalized as $ \int_{x_1}^{x_2} \psi_i(x) \phi_j(x)
dx=\delta_{ij} ,$ where $\delta_{ij}$ is the Kronecker delta ($\delta_{ij}=0$ for $i \neq
j$ and $\delta_{ij}=1$ for $i = j$).

With these relationships, we expand $P(x,t)$ in terms of these right eigenfunctions as
\begin{equation}
P(x,t)=\sum_{j=0}^{\infinity} a_j(t) \phi_j(x), \label{14}
\end{equation}
where the $\{a_i(t)\}$ are expansion coefficients that are related to the integration
\begin{equation} a_j(t)=\int_{x_1}^{x_2} \psi_j(x) P(x,t) dx. \label{10} \end{equation}

Next, we will express $\bar\mu(t)$ in terms of $\{a_i\}$ and $\{\lambda_i\}$.  From the
definition (\ref{33}) of $\bar\mu$, {
\begin{equation}
\bar{\mu}(t)=\int_{x_1}^{x_2} \mu(x) P(x,t) dx={\sum_{i=0}^{\infinity}} \lambda_i a_i(t)
\int_{x_1}^{x_2} \phi_i(x) dx\label{09},
\end{equation}
} where we have used the relation (\ref{02}), the boundary conditions (\ref{07}), and the
relations (\ref{13}) and (\ref{14}), successively.

The time evolution equation for $a_i(t)$ is straightforwardly obtained by inserting
(\ref{14}) into (\ref{39}), multiplying by $\psi_i(x)$ and integrating it over $x$:
\begin{equation}
\frac{d a_i(t)}{dt}=(\lambda_i-{\sum_{j=0}^{\infinity}} \lambda_j a_j(t) \int_{x_1}^{x_2}
\phi_i(x) dx )a_i(t) \label{11}
\end{equation}

In summary, the partial differential equation (\ref{06}) for $P$ is reduced to a set of
ordinary differential equations for $\{ a_i \}$, while the initial conditions of $a_i$ are
given from the relation (\ref{10}): $a_i(t_0)=\int_{x_1}^{x_2} \psi_i(x) P(x,t_0) dx$ for
the initial time $t_0$.

Note that there remains a freedom in the choice of $\phi_i(x)$ and $\psi_i(x)$, because
the normalization condition is still satisfied under the change of $\phi_i(x) \rightarrow
c_i \phi_i(x)$ and $\psi_i(x) \rightarrow (1/c_i)\psi_i(x)$ with any constant $c_i \neq
0$.  By taking advantage of this freedom, we can introduce, for convenience, another
normalization condition:
\begin{equation}
\int_{x_1}^{x_2} \phi_i(x) dx = 1 \label{08}
\end{equation}
for all the right eigenfunctions whose integral over $x$ does not vanish.  Indeed, this
normalization (\ref{08}) is easily achieved by re-scaling the eigenfunctions $\psi_i(x)
\rightarrow \psi_i(x) \int_{x_1}^{x_2} \phi_i(x')dx' $ and $\phi_i(x) \rightarrow
\phi_i(x)/\int_{x_1}^{x_2}\phi_i(x')dx'$.  If $\int_{x_1}^{x_2} \phi_i(x')dx'$ vanishes,
we simply leave the original eigenfunctions, and we call eigenfunctions with
$\int_{x_1}^{x_2} \phi_i(x')dx'=0$ "non-contributing eigenfunctions''.  Note that for the
0th right eigenfunction, $\phi_0$, this normalization is always possible, because the 0th
right eigenfunction does not take $\phi_0(x)=0$ for any $x$~\cite{kn:Morse}.  With this
choice of normalization, equation  (\ref{11}) is simplified as
\begin{equation}
\frac{d
a_i(t)}{dt}=(\lambda_i-{\sum_{j=0}^{\infinity}}' \lambda_j a_j(t)) a_i(t) \label{11b},
\end{equation}
where the prime over the summation symbol indicates that the summation is taken over all
eigenfunctions except non-contributing ones.

Equation (\ref{11b}) tells us that any eigenfunction $ \phi_i(x)$ of the linear operator
$L$, except for the non-contributing ones, gives a stationary solution of equation  (\ref{06}),
because any set \{ $a_i(t)=1$ and $a_j(t)=0$ for $j \neq i$ \} is a stationary solution of
(\ref{11}).  Among those stationary solutions, however, only the solution with
$a_j(t)=\delta_{j,0}$, is stable.

To show this, we make a linear stability analysis of these solutions.  Consider the
solution $a_i(t)=\delta_{ik}$ for given $k$, and introduce a perturbation $\delta a_i(t)$
as $a_i(t)=\delta_{ik}+\delta a_i(t)$ ($i=0,1,...,$).  Then, inserting this into
(\ref{11b}) and retaining only the terms of first order in $\delta a$, we obtain
$$\bibun{\delta a_i(t)}{t}=(\lambda_i-\lambda_k) \delta a_i(t) - \delta_{ik}
{\sum_{j=0}^{\infinity}}' \lambda_j \delta a_j(t) \equiv {\sum_{j=0}^{\infinity}}'
\Lambda_{ij}\delta a_j(t).$$

The eigenvalues of the matrix $\{ \Lambda_{ij} \}$ are easily shown to be
$(\lambda_0-\lambda_k)$, ... , $-\lambda_k$, $(\lambda_{k+1}-\lambda_k)$, ...  Recalling
that the eigenvalues are ordered so that $\lambda_i > \lambda_j$ for $i<j$, we can easily
show that all the stationary solutions for $k>0$ are unstable, while if $\lambda_0 > 0$
the solution with $k=0$ (i.e., with $a_i=\delta _{i0}$) is stable.  In other words, only
the mode with the largest growth rate remains as a stationary solution, as is expected.

The requirement $\lambda_0 > 0$ for the stability of the system is quite reasonable.
Otherwise, all $\lambda_j$ are negative, which means there is no growth at any state, and
all the cells would become extinct with time (recall that $\lambda_i$ is equal to the
growth rate of the mode represented by the $i$th eigenfunction).  To have a positive
growth rate for the stationary distribution, $\lambda_0 >0$ is therefore necessary.  The
condition $\lambda_0 > 0$ simply means that the cells (or units) continue reproduction
without extinction.

Now, the stationary solution of equation  (\ref{06}) is given by $\phi_0(x)$, the eigenfunction
of the operator $L$ corresponding to the maximal eigenvalue $\lambda_0$.  Similarly to the
case of the standard Fokker--Planck equation, the eigenvalue problem of the operator $L$
can be transformed into that for the Schr\"{o}dinger-type equation whose "potential'' is
given by the functions $f(x)$, $g(x)$, and $\mu(x)$ (see Appendix A).  Hence we can use
the methods and solutions developed in quantum mechanics.

\section{Two simple examples of the evolution of the distribution}

In this section we study two simple examples of equation  (\ref{06}) by linear or
threshold-type dependence of the growth rate on $x$.  We choose $f(x)=k x$ and $g(x)=D$ in
equation  (\ref{06}) with $k$ and $D$ positive constants; the reasons for this choice are: (i)
that the Gaussian distribution is often observed to be the stationary distribution of a
biological state, while this linear Langevin equation is the simplest to realize the
Gaussian distribution (the log-normal distribution is sometimes observed in cells
\cite{log-normal,Sato,Shiva}, but in this case we can simply use the logarithm of the
quantity as the variable $x$ that concerns us), and (ii) that this linear Langevin
equation has been thoroughly investigated in physics and mathematics; it models the motion
of a Brownian particle in a harmonic potential, so that we can easily see the effect of
the state-dependent growth introduced here.

\subsection{ $\mu(x)$ linearly dependent on $x$}

{ We study the case $\mu(x)=a x +b$ for $x$ to $[-\infinity, \infinity]$ in
equation  (\ref{06}), where $a$ and $b$ are constants. It is natural to study the linear case
as the simplest non-trivial example.  Indeed, as long as the range of $x$ in concern is
small, gradual change in $\mu(x)$ can be approximated by linear change.  }

In this case, we can obtain all eigenvalues and their corresponding eigenfunctions of $L$
as { $ \lambda_n = \frac{D a^2}{k^2}+b - k n $ and $\phi_n(x)=N_n H_n(
\sqrt{\frac{k}{2 D}} (x-\frac{2 D a}{k^2}) ) \exp [- \frac{k}{4 D} (x-\frac{2 D a}{k^2})^2
-\frac{k}{4 D} x^2] , $ } where $H_n(x)$ is the $n$th Hermite polynomial in $x$ and $N_n$
is the normalization constant determined by the normalization condition (\ref{08}).  In
particular, the stationary distribution is obtained directly as
\begin{equation} \phi_0(x)=N_0 \exp [-\frac{k}{2 D} (x-\frac{ D a}{k^2})^2 ]
\label{Gauss}, \end{equation} while the temporal evolution of the distribution is obtained
with these eigenvalues and eigenfunctions and with the reduced equations (\ref{11b}) for
$\{a_i\}$.

Fortunately, however, in this case there is a more convenient way to obtain the dynamics
of the system: if the system starts with a Gaussian distribution at some initial time, the
temporal evolution of the distribution preserves the Gaussian form.  By taking a Gaussian
distribution \begin{math} P(x,t)=\frac{1}{\sqrt{2 \pi \beta(t)}}
e^{-\frac{(x-\alpha(t))^2}{2 \beta(t)}} \end{math} with $\alpha$ and $\beta$ as the mean
value and the variance, it can be shown (see Appendix B), that the temporal evolution
preserves the Gaussian form when the time evolution equations for $\alpha$ and $\beta$ are
given by { $ \bibun{\alpha(t)}{t} = a \beta(t) - k \alpha(t) $ and $
\bibun{\beta(t)}{t} = - 2 k \beta(t) + 2 D .$ }

These equations indicate that while the temporal evolution of the variance is completely
the same as the case for a constant $\mu$, the evolution of the mean value is influenced
by the state-dependent growth; the mean value is shifted in the direction of larger $\mu$,
driven by its variance.  In the stationary state, as is also given in equation  (\ref{Gauss}),
the mean value (peak position) shifts with the degree $aD/k^2$ compared with the case
without the growth term (or, from the case with constant $\mu$ (i.e., $a=0$)).  Note that
this change in the mean value in the stationary state is proportional to the variance of
the original distribution, which is given by $D/k$, i.e.,
\begin{equation}
\Delta x=\frac{aD}{k^2}= \frac{a}{k} \Bra  (\delta x)^2 \Ket  \label{formula},
\end{equation}
where $\Bra ...\Ket $ is the average of the stationary distribution $P(x)$, and
$\delta x=x-\Bra x\Ket $.

In other words, the larger the variance of the distribution is, the more the mean value
shifts.  Correspondence with the fluctuation--response relationship \cite{Kubo,Sato} is
interesting, because the shift in the growth is proportional to the original fluctuation.
In addition, response to a higher growth state is possible only under the fluctuation of
the state, which demonstrates the relevance of phenotypic fluctuation to adaptation.  With
this shift of $\Delta x$, the average growth rate of a cell changes with
\begin{equation}
\Delta\overline{\mu}=a \Delta x ,
\label{delmu}
\end{equation}
which is an experimentally measurable quantity.  Hence, the right hand side of equation
(\ref{formula}) is represented by measurable quantities, because $k$ is simply the
relaxation time, $a$ is estimated from equation  (\ref{delmu}) and the variance $\Bra (\delta
x)^2\Ket $ is measurable.

\subsection{A threshold for growth: the step function $\mu(x)$}

{ We consider equation  (\ref{06}) with $\mu(x)= a, \; \Theta(x-x_0)+b$, where
$a$, $b$, and $x_0$ are constants, and $\Theta$ is the so-called Heaviside step function;
$\Theta(x)=0$ for $x < 0$ and $\Theta(x)=1$ for $x \geq 0$.  We study this case, because
in biological systems, a threshold for reproduction sometimes exists.  }

In this case, the eigenfunctions are written analytically with the use of confluent
geometric series and the corresponding eigenvalues are obtained, by transforming the
equation to the Schr\"odinger equation (see Appendix A).  Because the complete analytic
form is rather complicated, we discuss only the results of numerical calculations here.

\begin{figure}[hbtp] \begin{minipage}[t]{15cm} \begin{center}
\scalebox{0.2}{\includegraphics{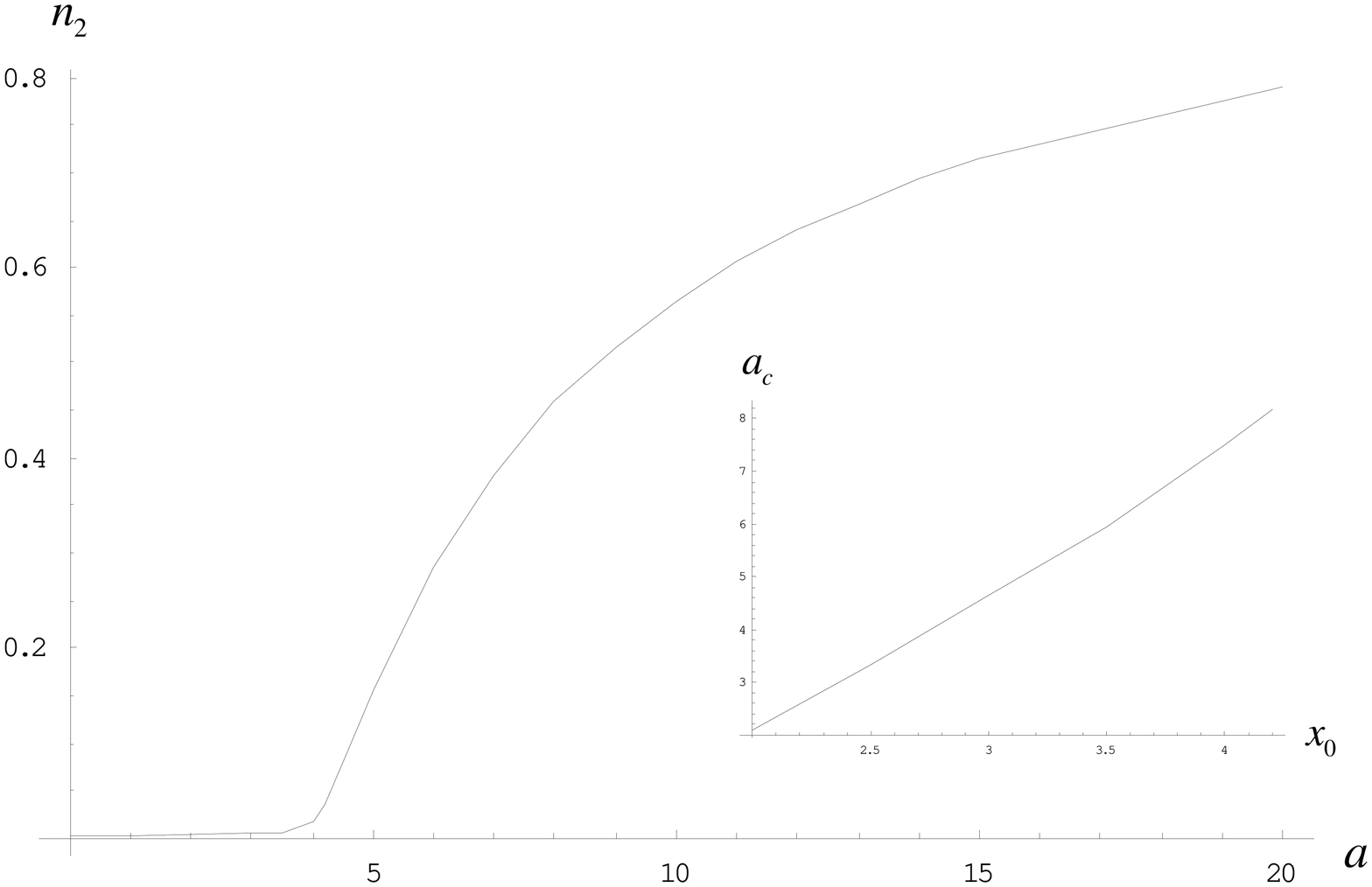}}
\caption{ The total amount $n_2$ of the distribution in the higher-growth region ($x> x_0
= 3$), calculated as $n_2=\int_{x_0}^{x_2} P(x) dx$, against the relative growth rate $a$.
This plot clearly indicates the existence of the critical value of $a$, $a_c$, which is
defined here as the value where $n_2=1/10$ .  The inset, a plot of the critical value of
$a_c$ against the position $x_0$ of the step of $\mu(x)$, shows the dependence of $a_c$ on
$x_0$, which is fitted well by: $a_c \simeq k x_0 \sqrt{\frac{k}{D}}$.  These calculations
were carried out for $f(x)=x$, $g(x)=1$, $\mu(x)=a \Theta(x-3)$, for the range of $x$,
$[-6,6]$.}
\label{34} \end{center} \end{minipage} \end{figure}

\begin{figure}[hbtp] \begin{minipage}[t]{15cm} \begin{center}
\scalebox{0.3}{\includegraphics{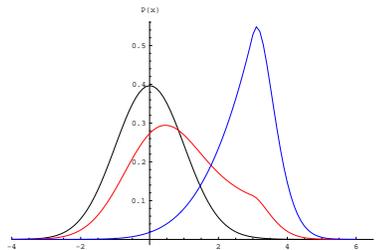}}
\caption{Some profiles of the stationary distributions for different relative growth rates
$a=2\mbox{(black)}, \; 5\mbox{(red)}$, and $8\mbox{(blue)}$.  We can see that the
distribution for $a=2$ is hardly influenced by $\mu(x)$.  We choose the same equation as
for Figure 1, i.e., $f(x)=x$, $g(x)=1$, $\mu(x)=a \Theta(x-3)$, with the range of $x$,
$[-6,6]$.}  \label{35} \end{center} \end{minipage}
\end{figure}

First, we consider the stationary distribution of equation (\ref{06}).  When the position
$x_0$ of the step of $\mu(x)$ is within the standard deviation of $P_{single}(x)$, i.e.,
$0 \leq x_0 < \sqrt{\frac{D}{k}}$ (we consider only the case of non-negative $x_0$), the
stationary distribution gradually moves toward the position $x_0$, as the parameter $a$
increases.  On the other hand, when the position $x_0$ is outside the standard deviation
of $P_{single}(x)$, i.e., $x_0 > \sqrt{\frac{D}{k}} $, the stationary distribution does
not change much until the parameter $a$ reaches some critical value $a_c$.  As $a$
increases beyond that value, the distribution shifts smoothly to larger $x$.  The
existence of the critical value $a_c$ is demonstrated in figure  (\ref{34}), which is a plot
of the total amount of the distribution in the right region ($x>x_0$) against the relative
growth rate $a$ (see figure (\ref{35})).

The critical value of $a_c$ is estimated to be $a_c \simeq k x_0 \sqrt{\frac{k}{D}} $, as
is confirmed numerically (see inset of Fig (\ref{34})).  Indeed, this value of $a_c$
coincides with the inverse of some characteristic time, that is the average time required
for a cell in a higher-growth state ($x>x_0$) to change to the lower-growth state ($ x <
x_0$).  This numerical result is reasonable: if the relative growth rate $a$ is smaller
than $a_c$, cells change to the state $x<x_0$ before they grow sufficiently in the
higher-growth region $x>x_0$.  The cells cannot `feel' the higher-growth region, so that
the difference in growth rates does not influence the cell population distribution.

\begin{figure}[hbtp] \begin{minipage}[t]{15cm} \begin{center}
\scalebox{0.3}{\includegraphics{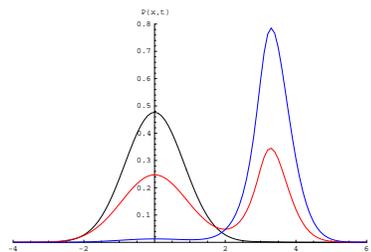}}
\caption{ Temporal evolution of the distribution for equation (\ref{06}) for $\mu(x)=20 \;
\Theta(x-3)$, $g(x)=1$, $f(x)=x$, with the range of $x$, $[-6,6]$.  The initial condition
is given by $P(x,t_0=0)=\delta(x)$.  The black, red, and blue curves show the
distributions at $t=0.6$, $1.04$, and $1.6$, respectively.  The double-peak distribution
is observed during an intermediate period.}  \label{36} \end{center} \end{minipage}
\end{figure}

Next, we briefly explain the dynamic behavior of the distribution when the relative growth
rate is greater than $a_c$ and the distribution is initially localized at $x<x_0$.  To be
specific, we set $P(x,t_0)=\delta(x)$, i.e., localized at $x=0$.  The temporal evolution
of the distribution is given in figure  (\ref{36}).  Here: (i) first, the distribution
behaves as if it does not `feel' the state-dependence of $\mu(x)$, until its tail touches
$x_0$, the edge of the step function.  (ii) After the tail of the distribution reaches the
edge of the step function, the distribution in this tail region starts to grow faster (see
figure  (\ref{36})); at this stage, the distribution has two peaks.  (iii) Finally, the
distribution converges to a single peak at the mean value at around $x_0$, the position of
the step of $\mu(x)$.  This temporal evolution to a higher growth state is in contrast to
the linear case, where a single-peak distribution is preserved and only the peak position
is shifted.

\begin{figure}[hbtp] \begin{minipage}[t]{15cm} \begin{center}
\scalebox{0.3}{\includegraphics{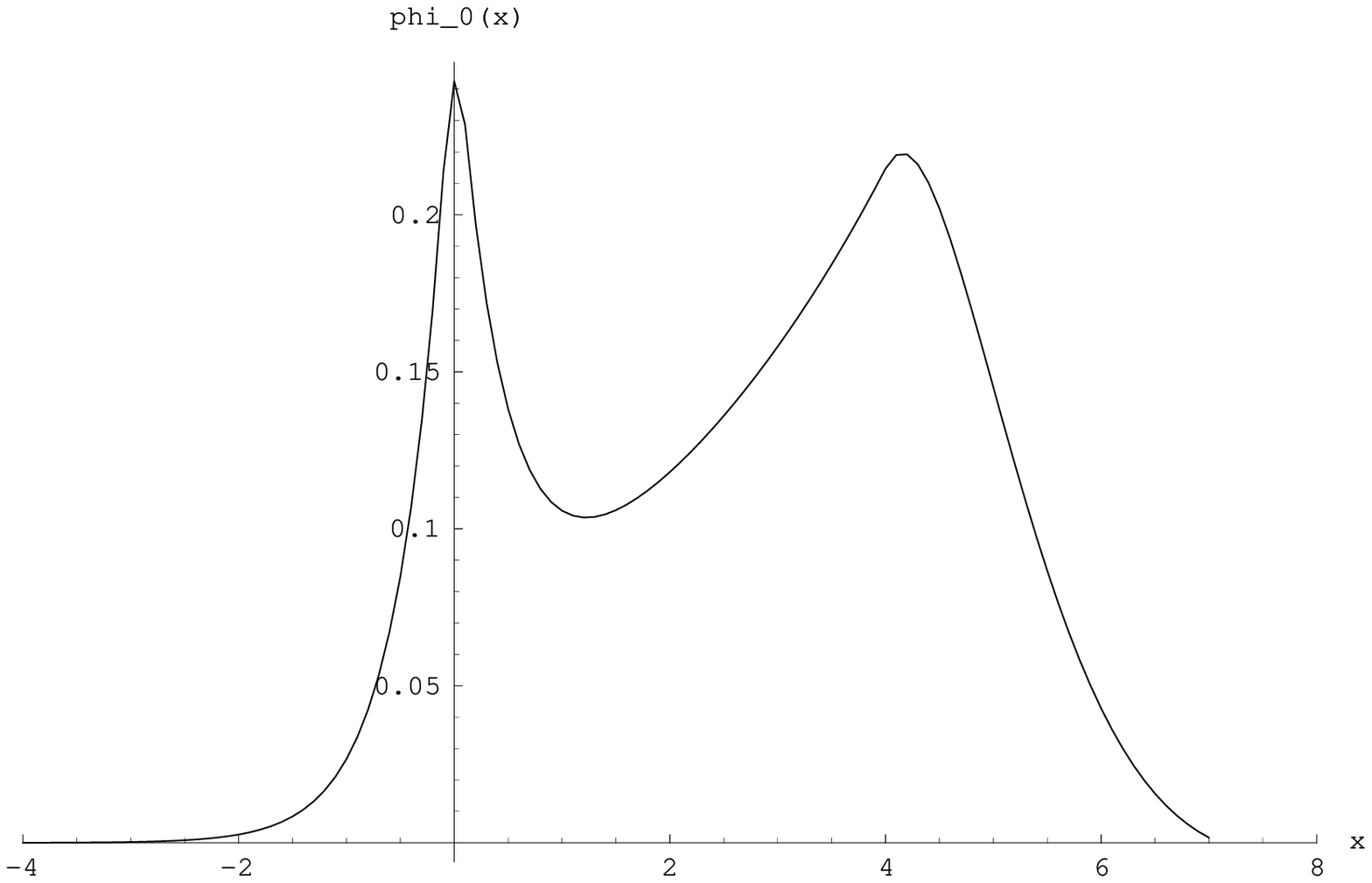}}
\caption{An example of the stationary distribution of equation (\ref{06}) having double
peaks, for $f(x)=2 \; \mbox{sgn}(x) $, $g(x)=1$, $\mu(x)=2.4 \; \Theta(x-4)$, with the
range of $x$, $[-7,7]$.}  \label{25}
\end{center} \end{minipage} \end{figure}

In the present example, the stationary distribution has a single peak.  For some forms of
$f(x)$, however, the stationary distribution has two peaks, even though the single cell
distribution (without the $x$ dependence of $\mu(x)$) has a single peak.  For example, for
$f(x)= 2 \mbox{sgn}(x)$ with the present form of $\mu(x)$, two peaks coexist (see figure
(\ref{25})).  Here, for large $x$ ($x>x_0(=4)$), the growth rate is high and the
distribution is confined within some range, so that the distribution has one peak in that
region, while for small $x$ ($x<x_0$), not all cells grow so that the distribution of the
cells tends to decrease.  However, many cells that have grown in the higher-growth region
flow into the lower-growth region because of the effect of the force of $f$, so that the
distribution has another peak there.

\section{
Conclusion and discussion }

In the present paper we have posed the question of how the distribution of an
intracellular state variable (say the abundances of some chemical or degree of gene
expression) is altered due to the state dependence of the replication rate of a cell.  To
discuss the temporal evolution of the distribution of the internal state $x$ of such
replication units, we have incorporated the state-dependent growth rate into the standard
Fokker--Planck equation.  By considering the population distribution of replication units
with Langevin equation dynamics, we have derived a general equation for the temporal
evolution of the distribution of states $P(x,t)$.  The derived equation includes a
self-consistent term arising from the growth rate.  In spite of this non-linear term, we
can formally solve the equation as an eigenvalue problem of the Sturm--Liouville type.
Note that the formalism presented here is rather general, as is the Fokker--Planck
equation.

After giving a general analysis of the equation, we have studied two simple examples,
assuming the linear Langevin equation for single-cellular dynamics.  First, when the
growth rate increases linearly with the state value $x$, the average of $x$ over cells
increases in proportion to its variance, which reminds us of the fluctuation-response
relationship in physics, while the proportion coefficient is estimated by the increase of
the growth rate and the relaxation time.  Note that the shift of population distribution
to a higher growth state is possible only with the fluctuation of the internal state.  Our
result implies that the response of $x$ to environmental change is proportional to its
variance.  In other words, fluctuations in chemical concentration, which have been studied
extensively, are relevant to biological adaptation.

Now let us return to the question raised in the introduction.  We measure an intracellular
state variable $\Bra x\Ket $, from an ensemble of cells, and study its change against the change
in external conditions.  Here we change the environmental condition (e.g., nutrient
concentration) and the cell state value $x$ (e.g., the concentration of some enzyme) is
changed accordingly.  After the cell distribution becomes stationary, we can measure this
change of the average $x$ denoted by $\Bra \Delta x\Ket _{total}$ that is caused by the change
in the environmental condition.  Now, from this measurement, we are often interested in
detecting the change in the stationary state of $x$, to explore intracellular dynamics.
However, such an intracellular state variable $x$ is often also related with the ability
for cell growth.  Hence $\Bra \Delta x\Ket _{total}$ is also influenced by the change in cell
growth speed, and this may deviate from the change caused by the intracellular dynamics
$\Bra \Delta x\Ket _{single}$.  Then, can we estimate the change of the internal state $\Bra \Delta
x\Ket _{single}$ from the measurement of $\Bra \Delta x\Ket _{total}$?  If we confine our
discussion only to the linear regime, we find \begin{equation} \Bra \Delta x\Ket _{total} =
\Bra \Delta x\Ket _{single}+\frac{a}{k}\Bra (\delta x)^2\Ket .  \end{equation} from equation
(\ref{formula}).  Here the latter term can be estimated from the standard measurements.
First, through equation (\ref{delmu}), $a$ can be estimated from the change in the average
growth rate of cells.  Second, $k$ is simply the relaxation time.  Hence, by measuring the
temporal change of $\Bra x(t)\Ket $, and by fitting the approach to its stationary value by an
exponential form, one can estimate $k$.  Finally, from the variance of the state value $x$
at a stationary state (by flow cytometry or by other means), we can obtain $\Bra (\delta
x)^2\Ket $.  Accordingly, we can estimate the term $\frac{a}{k}\Bra (\delta x)^2\Ket $, so that the
intracellular change of $x$ is estimated from the observable quantity $\Bra \Delta
x\Ket _{tot}$.

In our second example, we studied the case with a threshold-type dependence of the growth
rate on the state $x$.  When the position $x_0$ of the step of $\mu(x)$ is outside the
standard deviation of $P_{single}(x)$, i.e., when $x_0>\sqrt{\frac{D}{k}}$, the
distribution does not change significantly until the relative growth rate $a$ reaches a
critical value $a_c$, beyond which the distribution starts to shift to the higher-growth
region.  From the biophysical viewpoint, the value $a_c$ corresponds to the inverse of the
average time required for a cell to change from the higher-growth state ($x>x_0$) to the
lower-growth state($x<x_0$).

Here we have found that the distribution of the state variable often exhibits double peaks
over a long transient time.  For some form of $f(x)$ and $\mu(x)$, a double-peak
stationary distribution is also obtained, even if $P_{single}(x)$ has only a single peak.
This raises a cautious remark on the interpretation of the distribution observed in flow
cytometry.  Even if double peaks are observed, this does not necessarily mean that the
internal cell dynamics (e.g., gene expression network dynamics or metabolic dynamics) have
bistable states.  One of the peaks may be associated with the flow of population due to
the difference in reproduction speeds.

Several extensions of the present formulation are straightforward.  Although we mainly
discussed the case with a single state variable, extension to a higher-dimensional case is
straightforward.  Inclusion of a memory term to go beyond Markovian dynamics will be
possible, although we expect that most of the results on the linear and step-function
cases above are still valid in the non-Markovian case.

Although we have given our formulation here for a reproducing cell with an internal state
(e.g., chemical concentration), the present formulation can be applied generally to any
reproducing system with a growth rate dependent on its internal state.  For example, it
can be applied to an artificial cell or a replicating biochemical system with a growth
rate that depends on its internal catalytic activity.  Furthermore, application to
continuous evolution is possible.  By taking $x$ as a Hamming distance from a typical
gene, the evolution process to change $x$ to a given phenotype having some function can be
considered.  Here, the reproduction rate depends on $x$, which gives $\mu(x)$, while the
diffusion process in $x$ is simply the mutation, with $D$ as the mutation rate.  As
non-functional mutants are more common, the mutation in the change of function (or
activity) has a drift to a smaller regime, leading to a `force' term towards $x =0$ as in
equation(\ref{03}).  The temporal evolution of the distribution of gene $x$ is thus analyzed by
using our equation (\ref{06}), while in some examples, the steady state with positive
growth rate collapses~\cite{Kaneko}, with the increase of the mutation rate, as the
largest growth speed $\lambda_0$ becomes negative, which leads to error catastrophe.

A biological unit reproduces at a rate that depends on its state.  The present
Fokker--Planck equation with growth and death provides a basic equation for such problems
in general.

\ack 
We would like to thank T. Yomo and T. Suzuki for stimulating discussions.

\appendix

\section{Transformation of the linear operator $L$ to a Hermite operator}\label{19}

In this section we transform equation  (\ref{06}) to a type of Schr\"{o}dinger equation, to
show explicitly that the operator $L$ defined by (\ref{02}) is transformed to an Hermite
operator.  Here we follow the standard transformation from the Fokker--Planck equation to
the Schr\"{o}dinger equation~\cite{kn:Risken}, except for the existence of the terms
concerning $\mu(x)$.

We first introduce a new variable $y$ defined as $y(x)=\int_{x_0}^{x}
\sqrt{\frac{D}{g(x')}} dx'$, where $x_0$ is some number on $[x_1,x_2]$.  According to this
transformation, the distribution can change to $ \hat{P}(y,t)=\frac{1}{dy/dx}
P(x,t)=\sqrt{\frac{g(x)}{D}} P(x,t) $.  With these new variables, we can write equation
(\ref{06}) as:
\begin{equation}
\dot{\hat{P}}(y,t)=-\bar{\mu}(t)\hat{P}(y,t)+ \left[ \hat{\mu}(y)+\bubun{}{y} \left[
\hat{f}(y)+D \bubun{}{y} \right] \right] \hat{P}(y,t),
\label{32}
\end{equation}
where $\hat{f}(y)=\sqrt{\frac{D}{g(x)}}( f(x) +\frac{1}{2} g'(x) )$ and
$\hat{\mu}(y)=\mu(x(y))$.  $g'(x)$ is the derivative of $g$ with respect to $x$ and $x(y)$
is the inverse of the function $y(x)$.  Note that $\bar{\mu}$ does not change by this
transformation.

By further introducing two new quantities $\Phi(y)=\int_{y_0}^{y}\frac{\hat{f}(y')}{D}
dy'$ and $\Psi(y,t)=e^{\frac{\Phi(y)}{2}} \hat{P}(y,t)$, equation (\ref{32}) is rewritten
as
\begin{eqnarray}
\bubun{\Psi(y,t)}{t} \!\!\!&=&\!\!\!  -\bar{\mu}(t) \Psi(y,t)+\left[ V(y) + D
\bubun{{}^2}{y^2} \right] \Psi(y,t) \\ \!\!\!&=&\!\!\!  -\bar{\mu}(t) \Psi(y,t)+H(y)
\Psi(y,t)
\end{eqnarray}
where $ V(y)=\hat{\mu}(y)-\frac{\hat{f}(y)^2}{4 D}+\frac{\hat{f}'(y)}{2} $ and
$H(y)=\left[ V(y) + D \bubun{{}^2}{y^2} \right]$.  The operator $H$ obtained above is
evidently a Hermite operator, and indeed the eigenvalue problem of $H\Psi$ is simply a
type of Schr\"{o}dinger equation.  Accordingly, the exact solutions or techniques
developed for Schr\"{o}dinger equations can be applied to our problem.

\section{Temporal evolution preserving a Gaussian distribution for the linear $\mu(x)$ case}
\label{17}

When $f(x)=k x$, $g(x)=D$, and $\mu(x)=a x + b $, equation (\ref{06}) becomes
\begin{equation}
\bubun{P(x,t)}{t}=a( x - \Bra x \Ket _t)P(x,t)+\bubun{}{x} \left[ k x + D \bubun{}{x} \right]
P(x,t),\label{26}
\end{equation}
where we have used the normalization condition $\int_{x_1}^{x_2} P(x,t) dx=1$, and have
adopted the notation $\Bra ... \Ket _t \equiv \int_{x_1}^{x_2} ... P(x,t) dx$.  Multiplying both
sides of equation (\ref{26}) by $x$ and $x^2$ and integrating each case over $x$, we
obtain
\begin{eqnarray}
\frac{d \Bra x \Ket _t}{dt} \!\!\!&=&\!\!\!  a (\braket{x^2}_t - \braket{x}^2_t) - k
\braket{x}_t
\label{27}
\\ \frac{d \Bra x^2 \Ket _t}{dt} \!\!\!&=&\!\!\!  a (\braket{x^3}_t - \braket{x}_t
\braket{x^2}_t) - 2 k \braket{x^2}_t +2 D .
\label{28}
\end{eqnarray}

Suppose now that the solution of equation (\ref{26}) is a Gaussian distribution, i.e.,
\begin{equation}
P(x,t)=\frac{1}{\sqrt{2 \pi \beta(t)}} e^{-\frac{(x-\alpha(t))^2}{2 \beta(t)}} ,
\label{29}
\end{equation}
where $\alpha$ and $\beta$ correspond to the mean value of $x$ and its variance,
respectively, which are related to $\Bra x \Ket _t$ and $\braket{x^2}_t$ as $\alpha(t)=\Bra x \Ket _t$
and $\beta(t)= \braket{x^2}_t - \braket{x}^2_t$.  Using equations (\ref{27}) and
(\ref{28}) and the property of the Gaussian distribution $ \braket{x^3}_t=3 \alpha(t)
\beta(t)^2+ 3\alpha(t)^3 $, we can derive the time evolution equation of $\alpha$ and
$\beta$ as follows:
\begin{eqnarray}
\frac{d\alpha(t)}{dt} \!\!\!&=&\!\!\!  a \beta(t) - k \alpha(t)
\label{30}
\\ \frac{d\beta(t)}{dt}
\!\!\!&=&\!\!\!  \frac{d \Bra x^2 \Ket _t}{dt} -2 \braket{x}_t \frac{d \Bra x \Ket _t}{dt}
\nonumber
\\\!\!\!&=&\!\!\!  -2k \beta(t)+2D. 
\label{31}
\end{eqnarray} 

On the other hand, inserting the form of (\ref{29}) into equation (\ref{26}) and
simplifying the equation, we obtain the equation
$$ 2 (x-\alpha(t)) \beta(t) (k \alpha(t)-a \beta(t)+\frac{d\alpha(t)}{dt}) +
((x-\alpha(t))^2-\beta(t)) (-2D+ 2k \beta(t)+ \frac{d\beta(t)}{dt})=0. $$ The time
evolution equations of $\alpha$ and $\beta$ satisfy the above equations (\ref{30}) and
(\ref{31}), and the Gaussian distribution is the solution of equation (\ref{26}) (as the
solution with temporal evolution is unique).

\newpage
\noappendix

{\bf Figures}

\pagebreak


\section*{References}
\begin{thebibliography}{99}

\bibitem{Elowitz} Elowitz M B, Levine A J, Siggia E D and Swain P S 2002 {\sl Science}
     {\bf 297} 1183

\bibitem{Collins} Hasty J, Pradines J, Dolnik M, and Collins J J 2000 {\sl
    Proc.  Natl. Acad.  Sci. USA} {\bf 97} 2075

\bibitem{Ueda} Ueda M, Sako Y, Tanaka T, Devreotes P, and Yanagida T 2001 {\sl Science}
    {\bf 294} 864

\bibitem{Oosawa} Oosawa F 1975 {\sl J. Theor. Biol.} {\bf 52} 175

\bibitem{log-normal} Furusawa C, Suzuki T, Kashiwagi A, Yomo T and Kaneko K 2005 {\sl
    BIOPHYSICS} {\bf 1} 25

\bibitem{Siggia} Swain P S, Elowitz M B and Siggia E D 2002 {\sl Proc. Natl.  Acad.
    Sci. USA} {\bf 99} 12795

\bibitem{Elowitz2} Rosenfeld N, Young J W, Alon U, Swain P S, and Elowitz M B 2005 {\sl
    Science} {\bf 307} 1962

\bibitem{Kashiwagi} Kashiwagi A, Urabe I, Kaneko K and Yomo T.; submitted to {\sl Cell}.

\bibitem{Sato} Sato K, Ito Y, Yomo T and Kaneko K 2003 {\sl Proc. Nat. Acad. Sci. USA}
    {\bf 100} 14086

\bibitem{Kaneko} Kaneko K and Furusawa C 2005 {\sl J. Theo. Biol.} in press.

\bibitem{Arnold} Balagadde F K, You L, Hansen C L, Arnold F H and Quake
    S R 2005 {\sl Science} {\bf 309} 137

\bibitem{Haken} Haken H 1978 {\sl Synergetics: an introduction nonequilibrium phase
    transitions and self-organization in physics, chemistry and biology} 2nd edn
    (Springer-Verlag, Berlin) 

\bibitem{vanKampen} van Kampen N G 1992 {\sl Stochastic processes in physics and
    chemistry} (North-Holland, Amsterdam)

\bibitem{Paulsson} Paulsson J 2004 {\sl Nature} {\bf 427} 415

\bibitem{Arkin} Samoilov M, Plyasunov S and Arkin A P 2005 {\sl Proc.  Natl. Acad.
    Sci. USA} {\bf 102} 2310

\bibitem{Shibata-Fujimoto} Shibata T and Fujimoto K 2005 {\sl Proc.  Natl. Acad.
    Sci. USA} {\bf 102 } 331

\bibitem{kn:Risken} Risken H 1984 {\sl The Fokker-Planck equation; methods of solution
    and applications} (Springer-Verlag, Berlin)

\bibitem{kn:Morse} Morse P M, Feshbach H 1953 {\sl Methods of theoretical physics},
    (McGraw-Hill, New York) 

\bibitem{Shiva} Krishna S, Banerjee B, Ramakrishnan T V and Shivashankar G V 2005 {\sl
    Proc.  Natl. Acad.  Sci. USA} {\bf 102} 4771

\bibitem{Kubo} Kubo R, Toda M and Hashitsume N 1991 {\sl Nonequilibrium statistical
     mechanics } 2nd edn (Springer-Verlag, Berlin)

\end{thebibliography}
\end{document}